# A User Friendly Implementation for Efficiently Conducting Hammersmith Infant Neurological Examination

D. P. Dogra*, K. Nandam*, A. K. Majumdar*, S. Sural†, J. Mukhopadhyay*, B. Majumdar*
and
A. Singh‡, S. Mukherjee‡
*Department of Computer Sc. & Engineering, Indian Institute of Technology, Kharagpur, India
E-Mail: {dpdogra, nandam, akmj, jay }@cse.iitkgp.ernet.in, bandana.majumdar@gmail.com
†School of Information Technology, Indian Institute of Technology, Kharagpur, India
E-Mail: shamik@sit.iitkgp.ernet.in
‡Department of Neonatal Intensive Care Unit, IPGME & R and SSKM Hospital, Kolkata, India
E-Mail: drarunsingh61@yahoo.co.in, drsmukherjee70@gmail.com

*Abstract*—The aim of this work is to design a semi-automatic application that can be used as an aid by the doctors for smoothly conducting Hammersmith Infant Neurological Examination (HINE). A simplified version of the examination which provides a quantitative neurological assessment is used to design the application. The application includes a methodology of conducting HINE examination suited to inexperienced staff, applicable to both neonatal and post-neonatal infants. It also provides a facility to go through the previous records of a patient that can help in diagnosing patients with high risk of neurological disorder. A semi-automatic approach is proposed for skeleton generation. The application has been installed in hospitals and currently in operation. It is expected to increase the efficiency of conducting HINE using the proposed application.

I. INTRODUCTION

Intensive medical care performed in Neonatal Intensive Care Units (NICU) has increased the survival rate of very low birth weight and preterm newborns. An early prediction based on the outcome of various examinations conducted while the infant is admitted in NICU is clinically useful to counsel families who may benefit from early interventions [8]. A handful of methods have been developed in collaboration with scientists and medical professionals to minimize the error in such predictions. One such neurological examination method has been developed by Dubowitz et al. [2]. It has recently been updated for assessment of preterm and term infants to identify patients with higher risk of neurological abnormalities during later stage of their lives [3]. The Hammersmith Infant Neurological Examination (HINE) [2] [4] [11] is a part of their recommendation that uses the principles of standard neurological examination after the neonatal period. It is a simple and scorable method for assessing infants between 2 and 24 months of age which includes assessment of cranial nerve functions, posture, movements, tone and reflexes.

One purpose of HINE is to evaluate the neuromotor development of infants during first year of age and secondly, to correlate the scoring with levels of the Gross Motor Function Classication System (GMFCS) [7]. In particular, due to the high risk of Intra-Ventricular Haemorrhage (IVH) or Peri-Ventricular Leukomalacia (PVL), an increasing prevalence of Cerebral Palsy (CP) may occurred in premature, low birth-weight newborns and children born with asphyxia [4].

To the best of our knowledge, till date, medical professionals conduct HINE and record the outcomes manually on paper. No application is available that records visual evidence with the score while examinations are being conducted. Such an application can help to improve the assessment process of neuromotor development of a baby without compromising the quality of service. As software and video technologies has advanced rapidly during last decade, demands are now increasing to build software based on video to improve the performance of such services.

Several applications of health-care domain adopt computer vision based approaches. For example, an infant tele-monitoring system is proposed in [9] that uses safe, compact and non-invasive sensors to record the movements of a baby and use a client / server based approach for remote surveillance. A simulation based infant behaviour in a virtual environment is proposed in [6] where the authors try to avoid situations related to behavioural accidents. Though, a vision based approach can be regarded as a good alternative for conducting HINE, however, no such work is found in literature. This has motivated us to embed visual evidence to improve the efficacy of HINE. Issues that have been taken care of while designing the application, are listed as follows:

- We have designed an application for storing experiment outcomes including patient registration.
- Examinations conducted on both neonatal and post-neonatal babies are supported by the application.
- It is capable of capturing and saving videos of experiments that can be accessed in future while diagnosing a patient. For example, doctors can see the images / videos of any experiment that has been carried out on a baby at various gestational ages.
- It provides an interface to visualize templates of various examinations and corresponding video frames of the baby under examination.
- It adopts a semi-automatic approach. For example, while posture estimation is carried out, an auto generated skeleton is shown to the doctors such that they can easily compare it with reference templates.

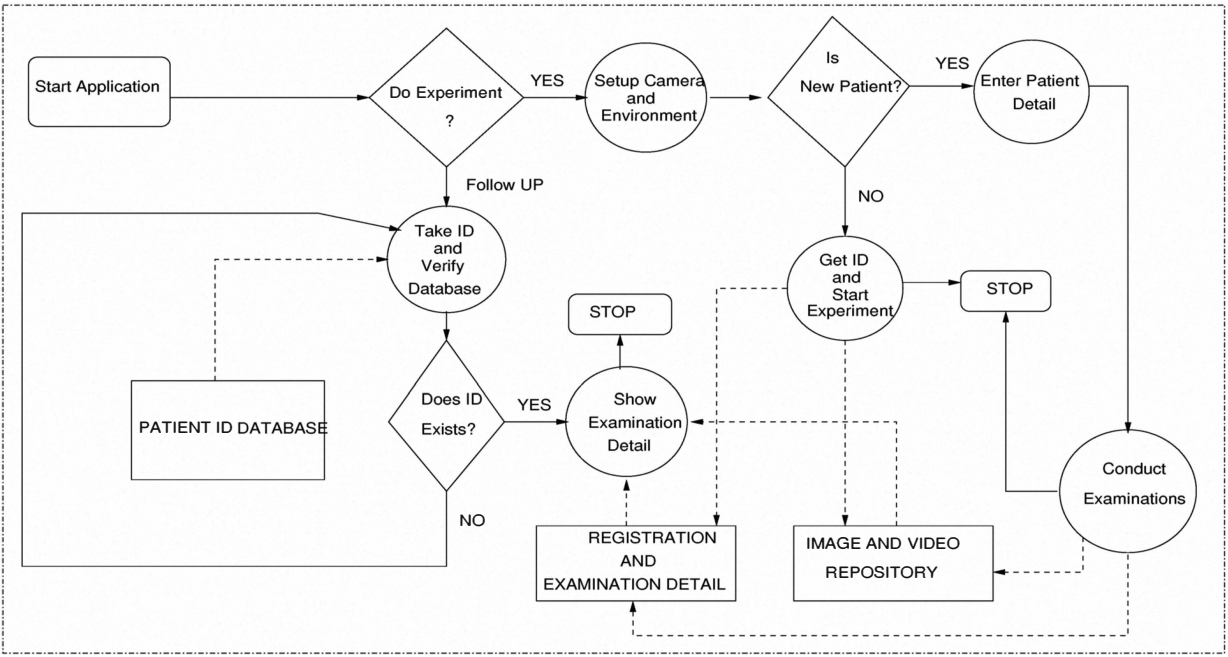

Fig. 1. A Flow Diagram of the Application for Conducting HINE for New and Follow Up Patients with Visual Recording.

The rest of the report is organized as follows. A detailed design of the proposed work including software design issues are outlined in Section II. Section III outlines usage of the application through screen shots with discussions highlighting a few points related to the future directions of this work. We conclude in Section IV.

## II. OVERVIEW OF THE PROPOSED WORK

To discuss about our scheme in detail, first, we present the overview of the proposed model using a flow diagram that is shown in Fig. 1. In following subsections, a detailed description about environmental setting, patient registration, examinations and follow-up management are described.

### A. Environmental Setup

A survey has been carried out at the examination site and a visual model for conducting HINE has been proposed. The arrangement shown in the model can capture best possible views of examinations. Observations suggest that if two cameras are fixed at predefined locations as depicted in Fig. 2, images can be captured with less intervention. Cameras are focused in a close shot view to minimize interference of external objects or movements. Both cameras support 25 frames / second (fps) video and CIF frame format (frame size 352 X 288 pixels). For acceptable visual recording, HINE is conducted in a controlled environment by maintaining similar illumination level throughout the entire period of examination. To process and save visuals of examinations, cameras are interfaced and controlled using a computer via video capture card.

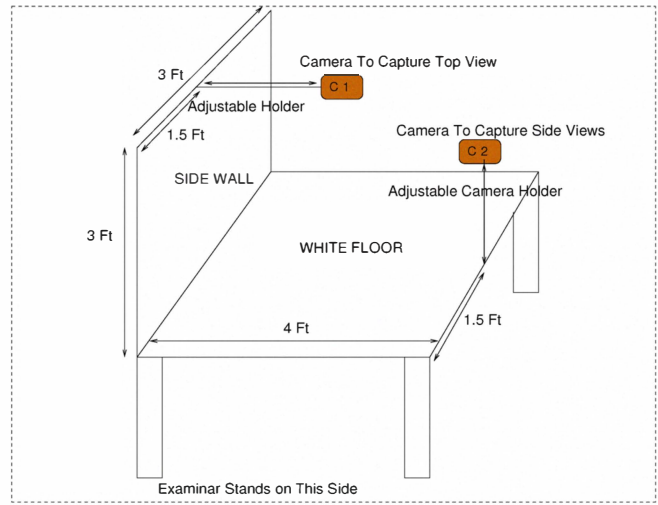

Fig. 2. Proposed Model for Recording Visual Evidence During HINE.

The examination desk is covered with soft mattress that is not considered as a potential health hazard. This is done in accordance with the well established setup of HINE in the Neonatal Intensive Care Unit (NICU) of Institute of Post-Graduate Medical Education & Research (IPGME & R) and Seth Sukhlal Karnani Memorial (SSKM) Hospital, Kolkata, India. In addition to that, to bring clarity in visual recordings, we have used white color coverings on both floor and side walls as shown in Fig. 2. As recommended by the doctors and medical professionals, clothing is removed from the baby including the head covering before examination commences. Temperature of the HINE laboratory is controlled by hospital authority such that newborn under consideration always remains in permissible environment.

*B. Patient Registration*

In this part of the discussion, patient registration process is described. After completing a survey and discussions with the doctors of NICU department of SSKM hospital, we have understood that the registration should cover two types of patients, i.e. (i) New Patient (ii) Old or Follow-Up Patient. In the former case, all information of a patient including name, date of birth, mother's name, father's name, gestational week at birth, corrected age, birth weight and discharge diagnosis are stored. The system automatically generates an unique patient identification number (ID) and the entire record is stored in a database. On the other hand, for follow-up patients, user is asked to provide the ID and a new entry is created in the database against it.

*C. Conducting Examinations*

In the NICU of SSKM hospital, both newborn and post-neonatal babies (upto 24 months) are considered for HINE when recommended by the doctors. Depending on the category of the infant, two sets of examinations recommended by the authors of [2] are used, one for neonates and other for post-neonatal infants. In the application under consideration, we have taken care of both types of patients. Ten examinations e.g. posture, arm recoil, arm traction, leg recoil, leg traction, popliteal angle, head control (extensor tone), head control (flexor tone), head lag and ventral suspension are carried out for neonatal patients. For every examination, 4 to 5 recommended models are shown on the screen and the examiner / doctor is asked to decide the state of the baby that best matches with one of these models. An example of posture estimation (carried out on a neonatal patient) is shown in Fig 3. For neonates, HINE is carried out only one time, at a gestational age of 40 weeks or above.

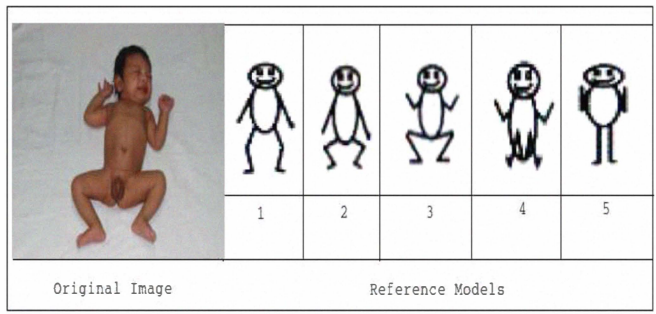

Fig. 3. Sample Image of Neonatal HINE. Examination: Posture Estimation.

For post-neonatal infants, a different set of examinations as recommended by the authors of [2] are carried out. It has three sections, namely, (i) Neurological items that include assessment of cranial nerve function, movements, tone, reflexes and reactions (ii) Motor milestones (iii) Behaviour. HINE is carried out at regular interval i.e., 3, 6, 9, 12, 15, 18, 21 and 24 months of age for post-neonatals. That is why, records of follow-up patients are managed differently. In future, data related to previous examinations are accessible to doctors for better diagnosis.

*D. Semi-automatic Approach*

A semi-automatic approach is adopted to aid in the process of estimating posture and other examinations. For the time being, an automatic skeleton extracting scheme, that is applicable only to the first examination of neonatals, is proposed. Let the image in Fig 3 be considered as a test case for automatic generation of posture. Once the image is captured using camera (C1), the process of automatic skeletonization is started. This is done as follows:

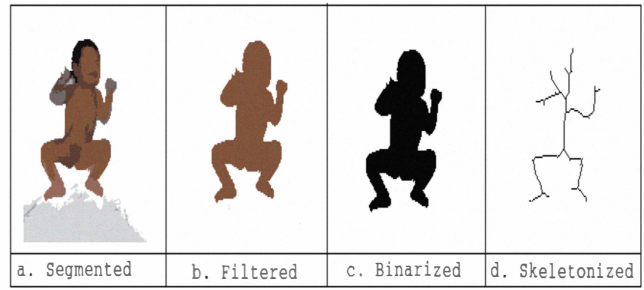

Fig. 4. Outputs at Various Steps While Producing Automatic Skeletonization of the Original Image Shown in Fig 3.

- To begin with, the image is segmented using HSV threshold based clustering proposed by Sural et al. [10]. This produces initial segments. Since, the background is homogeneous, segmentation is expected to produce an acceptable result (see Fig. 4a).
- We apply a region processing algorithm that merges small regions with large neighbouring regions. This step reduces the number of effective segments (see Fig. 4b).
- In the next step, foreground region with the largest contour size is assumed to be the expected infant silhouette. Background region is discarded (expected to be white). A binary image as shown in Fig. 4c, is produced.
- Finally, a thinning algorithm is applied to generate skeleton of the infant. An enhanced version of SPTA algorithm proposed by the authors of [5] produces acceptable results. In Fig. 4d, final skeleton of the image of Fig. 3 is shown.

*E. Follow Up Management*

Another feature of the application is to provide an interface for viewing previous examination scores of a baby. This is required for follow-up patients. Normally, a post-neonatal infant goes through a series of examinations at various gestational ages. To predict the neurological development of the baby in advance, doctors rely on viewing previous scores assigned to the baby. The application under consideration takes care of such issues. For example, at a given point of time, doctors can see the scores of an infant at previous examination timestamps. Similarly, videos and images corresponding to previous examinations can be fetched from the database.

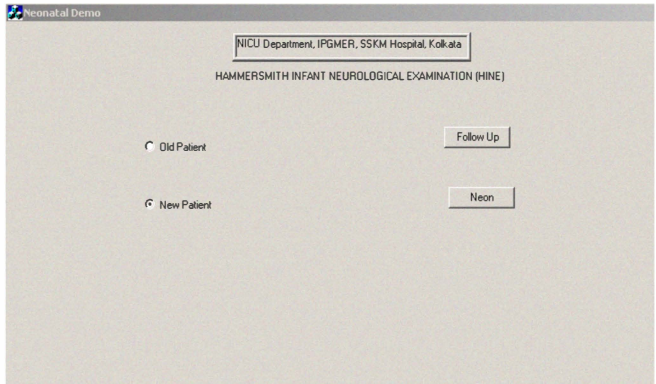

Fig. 5. A Screen Shot of the First Page of the Application.

## III. RESULTS AND DISCUSSIONS

The software is installed in the Neurological Examination Laboratory of the NICU department of IPGME & R, and SSKM hospital. An example screen shot of the starting window of the application is shown in Fig. 5.

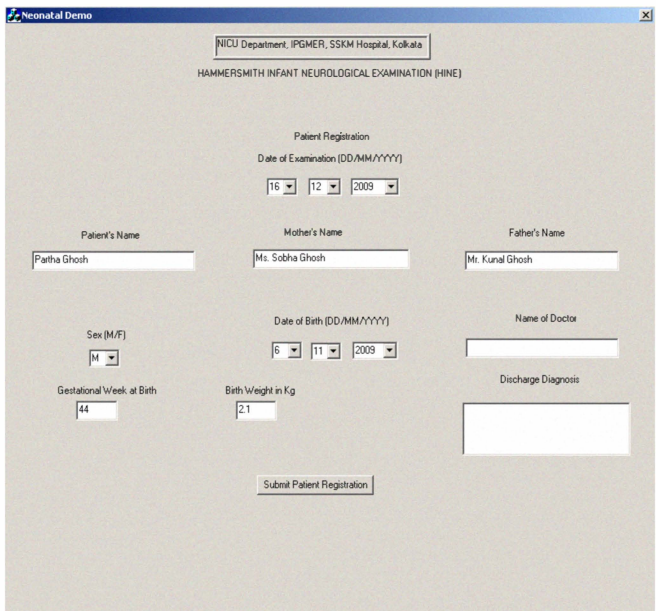

Fig. 6. Registration Page Used for a New Patient.

Through a set of images, intermediate states of the application while it is in use for recording measurements are shown. The image shown in Fig. 6 describes the registration page used for a new patient. Every important information that has been suggested by the doctors of SSKM hospital has been considered while designing the registration interface. Once it is submitted, the information is saved into a database and the system generates an unique patient ID that is delivered to the patient party for future references.

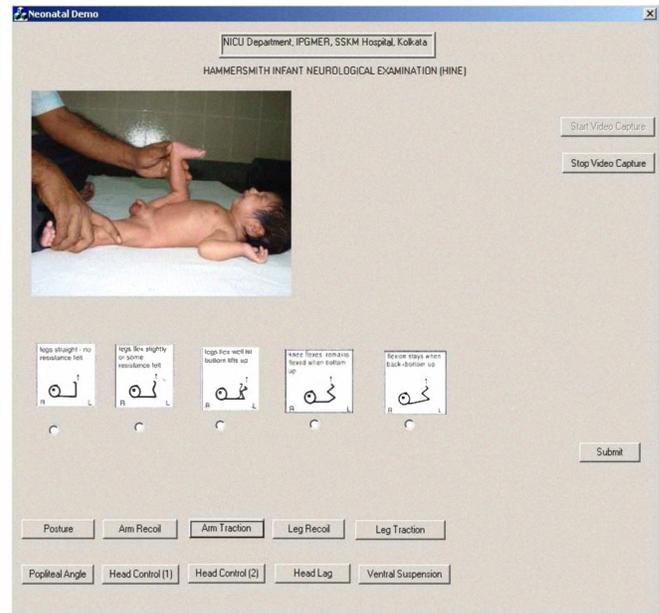

Fig. 7. An Example of a Leg-Traction Examination (neonatal patient).

The screen shot shown in Fig. 7, depicts an experimental environment while leg traction examination is being carried out (neonatal patient). At the same time, a video frame that shows the state of the infant is embedded on the screen. Concurrently, it is saved into the database for future references. In the later phase of this work, we aim to represent the scene using visual models.

A semi-automatic scheme to generate skeleton of the infant during posture estimation is embedded into the current application. In Fig. 8, while the posture estimation is being carried out, simultaneously, skeleton of the baby is extracted and embedded on the screen. This helps the doctors for making better assessment.

### A. Future Plans

There are a few limitations of the current application that are being seriously considered to be addressed in future.

- There are plans to design automatic feature generation for other examinations of the Hammersmith chart using computer vision based algorithms.
- Association of the outputs with predefined templates such that the system can automatically generate score, is also in the pipeline of our future work. For example, a scheme to associate the output skeleton (as depicted in Fig. 8) with the template that matches best, is seriously in consideration. A graph based algorithm can be used.

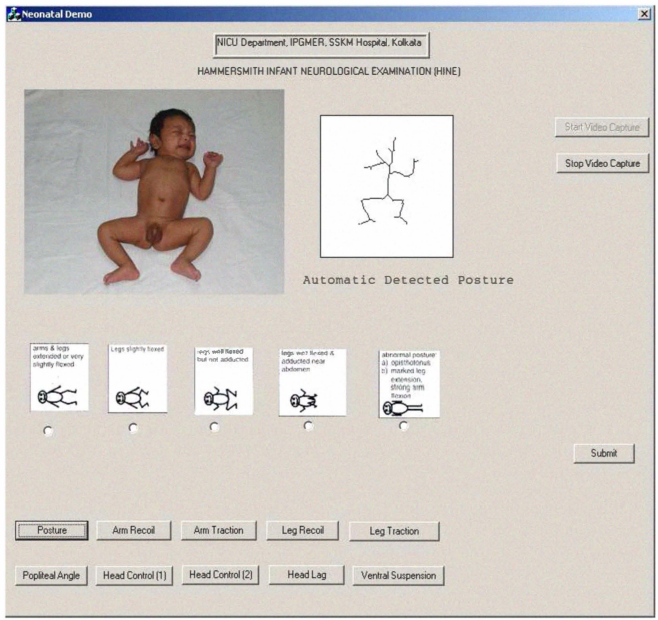

Fig. 8. An Example of Posture Detection Examination (neonatal patient).

- Once, a considerable amount of data is collected from hospitals, techniques like data mining can be used for automatic prediction of the neurological development index of a baby.

## IV. CONCLUSION

Hammersmith Infant Neurological Examination is a popular method to assess the neurological development of infants, especially for newborns between 2 to 24 months of age. Evidence found by the medical researchers that preterm infants are likely to go through neurological disorder during later phases of life. An early detection of such disorder is helpful in diagnosing such abnormalities. Doctors use HINE to detect such cases. To aid in the process of conducting HINE, an efficient approach is proposed. We have developed an application that is capable of recording examination details, patient information and visual evidences. A semi automatic design is proposed for posture estimation of the neonatal as mentioned in the Hammersmith chart. This can be used to reduce the workload of the persons involved in such environment. It is expected that, once it is made fully functional as mentioned in the future plans, this will be a considerable aid to the NICUs.

## ACKNOWLEDGEMENT

We acknowledge the support provided by Department of Information Technology, Ministry of Communication and Information Technology, Govt. of India for funding this project. We are also thankful to NICU department of IPGME & R and SSKM Hospital, Kolkata for necessary support. We would like to thank WEBEL Corporation Ltd., Kolkata, India for their help in carrying out the project smoothly.